# Nonlinearity in a Crosslinked Polyelectric Polypeptide

Jorge Monreal[a], John Schreiber[a], Donald T. Haynie[a,b]

[a]Department of Physics, University of South Florida, 4202 East Fowler Avenue, Tampa, FL 33620, USA

[b]Nanomedicine and Nanobiotechnology Laboratory, Department of Physics, University of South Florida, 4202 East Fowler Avenue, Tampa, FL 33620, USA

Correspondence to: jmonreal@alum.mit.edu (Jorge Monreal)

**ABSTRACT**

Young's modulus of soft solids composed of crosslinked synthetic polypeptides has been determined under different conditions. Co-poly-(L-glutamic acid$_4$, L-tyrosine$_1$) [PLEY (4:1)] was crosslinked with poly-L-lysine (PLK) and 1-ethyl-3-(3-dimethylaminopropyl) carbodiimide hydrochloride (EDC). Elasticity was assessed by subjecting samples to a compressive strain. Crosslinked material at high relative humidity, RH 75-85%, exhibited non-linear elasticity. Stress-strain response was approximately linear at low strain but nonlinear above a threshold strain. Analysis of the secant modulus revealed apparent softening of samples at low strain and hardening at high strain, as in biological soft tissues. Fitting stress-strain data with a neo-Hookean model yielded approximately $40 \leq E \leq 300$ kPa at high RH. Viscoelasticity was nonlinear at low RH. The average viscosity-driven relaxation time was 13 min at high strain and 6 min at low strain. Analysis of the derivative of the secant modulus for non-linear elastic materials revealed a transient response up to a strain of $\varepsilon \approx 0.18\text{-}0.20$. Above this range, oscillations tended to zero. Non-linear viscoelastic materials showed lower-amplitude oscillations than samples at high RH up to ε ≈ 0.06 and strong damping thereafter. The data suggest that it will be possible to engineer mechanical properties of polypeptide materials.

**INTRODUCTION**

Protein- and polypeptide-based biomaterials are of increasing interest in medicine, biotechnology and biodegradable materials.[1,2] Advantages of structural proteins for such materials include intrinsic biocompatibility, the ability to self-assemble into complex higher-order structures, for example, collagen fibers, and a remarkable range of mechanical properties, for example, high-performance elasticity and toughness.[1,2] Some protein elastomers can withstand over 100% elongation without rupture and return to the original length on removal of stress.[3,4] The rubber-likeness displayed by some proteins will depend on physical properties of individual chains.

Most structural proteins have repetitive amino acid sequences.[5,6] Different sequence motifs are found in different structural proteins, which display different mechanical properties and biological functions.[5,6] Specifically how amino acid sequence translates into protein elasticity is, however, largely unclear. The essential features of rubber-like elasticity are, by contrast, clear enough: long chains enable deformation, at least some independence in chain behavior is required, and crosslinks limit deformation.[7] One of the most extensively studied elastic proteins is the connective tissue protein elastin. The wild-type protein features numerous valine-proline-glycine-valine-glycine repeats, and lysine residues enable enzyme-catalyzed crosslinking of chains, turning soluble individual chains into an insoluble fibrous aggregate.[8-10]

There are three main models of elastic elasticity: random chain network, a solvent-related mechanism and extension-dependent damping of internal chain dynamics. The random chain network model was developed by Flory.[11,12] Hoeve and Flory reported a low value the ratio of the internal energy to the total elastic force for bovine elastin and concluded that the material was a network of random chains.[13] The authors claimed the viewpoint was affirmed over a decade later.[14] In the model of Weis-Fogh and Anderson, by contrast, extending an elastin fiber will increase the exposure of hydrophobic side chains to solvent, lowering the entropy of water as it forms a cage at the fiber-solvent interface.[15-17] The backbone of a polymer, however, not the solvent, must bear the tensile load. As to chain dynamics, the β spiral conformation of an elastin chain will permit oscillations of the ϕ and ψ angles of amino acid residues other than proline. Chain stretching could dampen the amplitude and influence the frequency of the oscillations.[18,19] This view appears to gain support from dielectric relaxation and acoustic absorption experiments and computational studies.[20] The amplitude of the oscillations will, however, depend mainly on thermal energy; a compressive or tensile force is not expected to have marked impact on the oscillations, provided that the chain is not stretched too much. The elastic properties of other proteins are assumed to depend on different mechanisms, though of course basic principles will be generally valid.

Crosslinked polymer networks display non-linear elasticity under some conditions. Various theoretical and descriptive models have been proposed. Rubinstein and Panyukov, for example, have developed a molecular model of non-linear elasticity for entangled polymer networks.[21] Storm et al. have proposed a molecular model of the non-linear elasticity of actin, collagen, fibrin, vimentin and neuro filaments.[22] More recently, Carrillo et al. used a combination of theoretical analysis and molecular dynamics simulations to develop a model of networked deformations, which the authors then used to describe non-linear mechanical properties of polymers and biological gel networks.[23] Synthetic polymer networks deform reversibly at an applied stress in the $10^4$–$10^7$ Pa range. Networks of the proteins actin and collagen, by contrast, deform at stresses as low as $10^{-1}$–$10^2$ Pa. The elastic response of a polymer network will normally contain both entropic and enthalpic components, the balance depending on the polymers involved and how they interact.

Here we assayed for the response of crosslinked synthetic polypeptides to an applied compressive stress. The polymer chosen for the experiments was a random co-polymer of L-glutamic acid (E) and L-tyrosine (Y) in a four-to-one molar ratio [PLEY (4:1)]. PLEY molecules were crosslinked with PLK, a polycationic homopolymer, and EDC, a diimide reagent. We then compared the mechanical properties of the materials with biological tissues, polymer networks, and biological gels. To the authors' knowledge, there are no available studies on cross-linked PLEY (4:1) in the field. However, synthetically designed polypeptides are becoming increasingly important in diverse areas such as biodegradable devices, medical implants and mechanical dampers. This study initiates mechanical property studies of materials that have not previously been investigated. The data suggest that it is possible to engineer mechanical properties of polypeptide materials.

**EXPERIMENTAL**

**Materials and Methods**

*Polymers*

PLEY (4:1) (MW 20-50 kDa) and PLK (MW 15-30 kDa) were from Sigma-Aldrich (USA). EDC was obtained from TCI America (USA). The chemical structures of the side chains are shown in Figure 1. The nominal $pK_a$s are 4.1 (E), 10.5 (Y) and 10.5 (K).

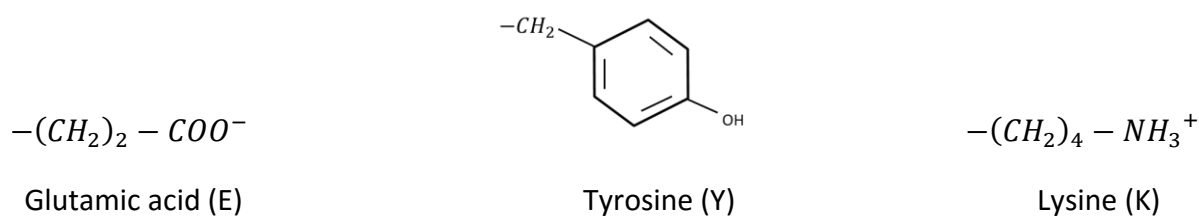

$-(CH_2)_2 - COO^-$           Tyrosine (Y)          $-(CH_2)_4 - NH_3^+$

Glutamic acid (E)                   Lysine (K)

**Figure 1.** *Side chains of the amino acids of the present study. The side chain of an amino acid residue is attached to an alpha carbon atom in the polymer backbone. A peptide group is formed between backbone atoms of successive amino acid monomers.*

*Mold*

Samples were formed in a custom-made aluminum mold. Fabricated in the USF Department of Physics machine shop, the base and lid of the mold were cut from a 0.5''-thick plate, cylinders were cut into a 0.125''-thick plate, and all flat surfaces were milled to a roughness of 2.5-25 μm. The mold was designed to produce 1 to 1, length to diameter cylinders, each having a volume of ≈50 mm³. All samples had a nominal diameter and height of 4 mm.

**Table 1:** *Experimental factors of this study.*

| Factor | Level | | |
|---|---|---|---|
| | -1 | 0 | +1 |
| PLEY (4:1) concentration (% w/v) | 30 | 40 | 50 |
| PLK concentration (% w/v) | 10 | 15 | 20 |
| EDC concentration (% w/v) | 20 (~1.0 M) | 30 (~1.5 M) | 40 (~2 M) |
| Relative Humidity (%) | 33 | 75 | 85 |

*Experimental Design*

Table 1 summarizes the details of sample composition. Three experimental factors were the concentrations of PLEY (4:1), PLK and EDC, and each had three possible values. RH, a fourth factor, also had three possible values. A full-factorial experiment would have involved $3^4$ = 81 trials, exceeding the scope of the project, so the number of trials was reduced to $3^3$ = 27. Table S1 in Supplementary Material, generated with JMP statistical software (SAS, USA), shows a randomized trial for the selected factors. PLK at low concentration was predicted to be unlikely to precipitate PLEY (4:1) by way of interpolyelectrolyte complex formation, the lysine side chains of PLK binding to glutamic acid side chains of PLEY by Coulomb interactions. The three-dimensional polymer network thus formed is more likely to precipitate at a high concentration of PLK, as the odds of saturating available sites on PLEY will be higher. Table S1 also presents the composition of each of the 27 samples of this study. Sample 1, for instance, was prepared by mixing 40 µL of 30% (w/v) PLEY with 40 µL of 40% (w/v) EDC and 40 µL of 20% (w/v) PLK. The sample was then equilibrated in a chamber at 85% relative humidity prior to analysis. The sample compositions of this study were 3 replicates of 9 different conditions, not 27 different conditions. Not all samples survived processing. This is evident in conditions with only two replicates as presented in raw data graphs Figures 3-6. Materials were fabricated down the list according to Table S1 so as to not introduce bias.

*Materials Fabrication*

Equal volumes (40 µL) of PLEY solution, EDC solution and PLK solution were mixed in a 1.5-mL tube using a 250 µL micro-pipette. Volumes delivered were expected to be within 2 µL of nominal, or 5% variation. EDC was mixed first with PLEY to prevent immediate crosslinking. EDC is reactive towards carboxylic acid, but the anhydride is unstable; a peptide bond is formed when EDC reacts with a carboxylic acid group and an amino group. PLK was then added to the PLEY-EDC mixture. Nominal concentrations at 30%, 40% and 50% PLEY were made thus. Unfortunately, there are no chromophores available to determine reactant concentrations more accurately with UV instruments. The final reaction mixtures were then transferred to the mold with a 1-mL syringe and allowed to set up over a 24-h period. Samples were removed from the mold and placed in a humidity chamber. Each sample was allowed to come to equilibrium over a period of 3 days prior to mechanical analysis. The samples had a diameter of (3.5 ± 0.4) mm and a height of (3.6 ± 0.3) mm. Individual height and diameter measurements were considered accurate to within 0.5 mm.

*Humidity Control*

Desired RH values were obtained with different saturated aqueous solutions of salt: KCl for 85%, NaCl for 75% and MgCl$_2$ for 33%.[24] Saturated salt solutions were prepared by adding deionized water to a minimum of 5 g of salt previously deposited in a 15-mm tall petri dish. Saturated salt solutions were deposited in separate petri dishes, sealed with parafilm and allowed to come to equilibrium at 22 °C over several days. This method has been shown to provide relative humidities with 2% of nominal values, if prepared correctly. Samples were removed from the mold and placed in the designated relative humidity chamber. Each sample was then allowed to come to equilibrium at 22 °C over a period of 3 days prior to mechanical analysis. Figure S1 in supporting information provides experimental evidence to support the chosen equilibrium period.

*Materials Testing*

*Force-Displacement (F-D) measurements*

Uniaxial compression tests of crosslinked samples were performed as illustrated in Figure 2. A digital scale (Centech, USA) with a maximum loading capacity of 1000 g (9.8 N) and a resolution of 0.1 g was placed on top of a lab jack, giving continuous displacement in the vertical dimension. An analog displacement gauge (Pittsburgh Dial Indicator, USA, 0.001 in. (0.025 mm) resolution) was used to quantify compression. The sample was placed between the scale and a fixed arm. This system enabled measurement of spring constants in the 10-1100 N/m range. Displacement was the independent variable; compressive force was the dependent variable. A loading curve was obtained for each sample, an unloading curve for selected samples. All data were acquired at 22 °C. Figure S2 in supporting information shows the loading and unloading curve of the apparatus itself over the force range of 0.5-6 N.

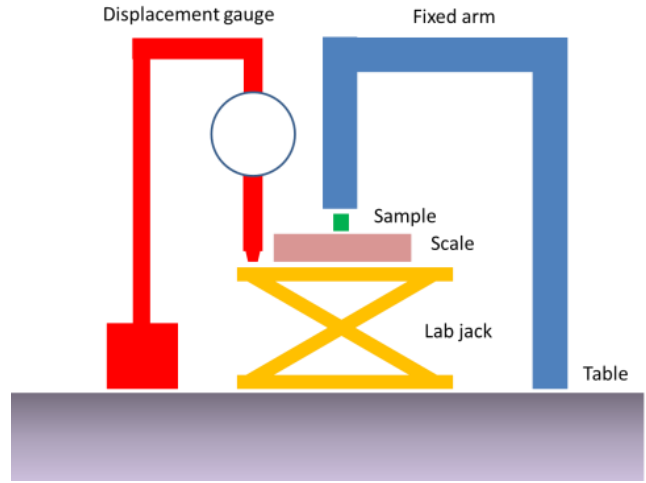

*Figure 2. Measurement apparatus. Cylindrical samples were compressed uniaxially along the vertical axis by adjusting the height of the lab jack. Displacement was quantified with a displacement gauge.*

*Relaxation time measurements*

Conditions at 33% relative humidity were analyzed through relaxation times. To obtain relaxation times, a sample was placed on a lab scale utilizing the apparatus shown in Figure 2. A predetermined strain was applied to the sample by raising the lab jack by a displacement coinciding with the desired strain. At the point where the desired strain was reached, as measured by the displacement gauge, the sample was allowed to relax down to its equilibrium value over time. Values of stress were determined by the values of weight measured on the balance. To capture relaxation as a function of time, a video of the relaxation process was started at the point of maximum stress for the given strain. Relaxation stress values at each time were extracted from analysis of the video.

***Fitting and Parameter Determination***

*Young's Modulus*

There are several phenomenological and mechanistic models of varying levels of complexity that can be used to describe hyperelastic materials. For biological materials and cross-linked polymers, specifically, a neo-Hookean model is widely used. There are some disadvantages to the use of a neo-Hookean model. In general, its predictive nature is less accurate than models such as the Mooney-Rivlin, Ogden and Arruda-Boyce models, as examples.[25,26] Additionally, the neo-Hookean model does not predict accurately at large strains. However, given that our samples showed significant variability it was decided that the neo-Hookean model is appropriate to serve our purpose of elasticity comparisons among different conditions. The simplicity of the model allowed for a standard method to be applied to each condition and, thereby, determine if patterns arise due to changing experimental conditions of concentration and relative humidity. This study did not seek to determine exact values of shear moduli at each given condition. Rather, our aim was to understand changes in elasticity due to changes in condition on a condition-to-condition comparative basis. We, nonetheless, compared stiffness trends obtained by the neo-Hookean model with that of a two-parameter Mooney-Rivlin model.

All fitting parameters and confidence intervals were obtained with Matlab®, implementing a trust region nonlinear least squares algorithm.

<u>Young's modulus</u>

Engineering normal stress is calculated as $\sigma = F/A_o$, where F is the applied force along a single axis and $A_0$ is the original cross-sectional area of the sample. Engineering strain is defined as $\varepsilon = \frac{l - L_o}{L_o}$, where $l = L_o - \delta$ is the length of the sample after a compressive displacement $\delta$ and $L_o$ is the original sample length. Principal stretches are defined by engineering strain, $\varepsilon = \lambda - 1$, where $\lambda = \frac{l}{L_o}$. For an incompressible, homogeneous and isotropic material, true stress $\sigma_{true} = \mu\left(\lambda^2 - \frac{1}{\lambda}\right)$, where $\mu = \frac{E}{2(1+\nu)}$ is the shear modulus, $\nu$ is Poisson's ratio, and $E$ is Young's modulus. The engineering normal stress will be $\sigma_{eng} = \mu\left(\lambda - \frac{1}{\lambda^2}\right)$ on inserting $\sigma_{eng} = \frac{\sigma_{true}}{(1+\varepsilon)}$. If $\nu = 0.5$, as is the case for incompressible isotropic materials undergoing isochoric deformation, then $E = 3\mu$. The neo-Hookean stress model for engineering stress can be fit to stress-strain data to determine $\mu$ and thus $E$.

The Mooney-Rivlin model used here contains two parameters, though up to nine parameters can be used. These material constants, as they are called, are $C_1$ and $C_2$ and here expressed as $\mu_1 = 2C_1$ and $\mu_2 = 2C_2$ to keep notation comparable to that used in the neo-Hookean model. The Mooney-Rivlin model is in essence an extension to the neo-Hookean model. The engineering normal stress will be $\sigma_{eng} = \left(\mu_1 + \frac{\mu_2}{\lambda}\right)\left(\lambda - \frac{1}{\lambda^2}\right)$. Linearization of this equation gives the form $\sigma_{eng}* = \mu_1 + \mu_2\beta$, where $\sigma_{eng}* = \sigma_{eng}/\left(\lambda - \frac{1}{\lambda^2}\right)$ and $\beta = 1/\lambda$. On a plot of $\sigma_{eng}*$ versus $\beta$ one finds that $\mu_1$ is the slope. In general, due to our compressive tests we will find negative values of $\mu_2$.

Relaxation time

The time dependence of stress at a given strain was modeled as follows. For two elements of a Maxwell material in parallel, each consisting of a spring and a dashpot in series, the differential equation for the stress-strain response is $\sigma + p_1\dot{\sigma} + p_2\ddot{\sigma} = q_1\dot{\varepsilon} + q_2\ddot{\varepsilon}$, where $p_1 = \tau_1 + \tau_2$; $p_2 = \tau_1\tau_2$; $q_1 = \eta_1 + \eta_2$; $q_2 = \eta_1\tau_2 + \tau_1\eta_2$. Here $\tau_i = \frac{\eta_i}{E_i}$, $i \in \{1,2\}$ is relaxation time and $\eta_i$ is viscosity. Relaxation times were determined by fitting $\sigma(t) = \varepsilon_o \left(E_1 e^{-\frac{t}{\tau_1}} + E_2 e^{-\frac{t}{\tau_2}}\right)$ to individual experimental data sets.

**RESULTS AND DISCUSSION**

*Force Versus Displacement*

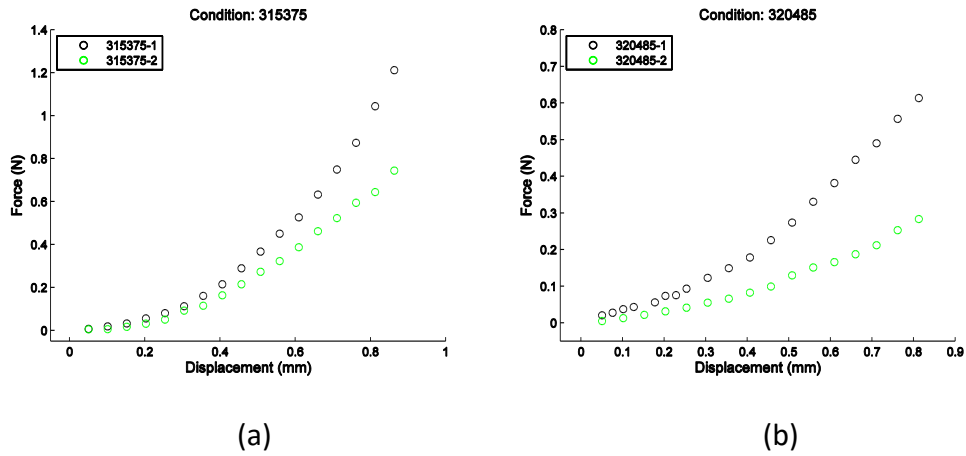

(a)        (b)

*Figure 3: Force-displacement loading curves of conditions at a.) 75% and b.) 85% relative humidities at 30% w/v nominal PLEY concentration.*

A force-displacement curve was obtained for each sample. Figures 3-5 show loading curves for samples in each of the conditions at relative humidities of 75% and 85%. See Table S1 in Supplementary Material for specific compositions and summary of samples within each condition. Labels for each condition contain information about PLEY, PLK and EDC concentration as well as relative humidity. For example, 315375 represents 30% (w/v) PLEY, 15% (w/v) PLK, 30% (w/v) EDC and 75% RH. Not all samples survived processing. This is evident in conditions with only two replicates. We, additionally, obtained force-displacement curves for samples at relative humidities of 33%. These are shown in Figures 6a-6c. Figure 7 presents force-displacement data for all conditions with each condition averaged over replicates. To obtain the error bars in the force-displacement plot the standard deviation in weight among samples for each condition was obtained at each displacement value setting. Error in force was calculated as $F = C(w \pm \Delta\varepsilon_w)$, where $w$ is weight and $\Delta\varepsilon_w$ is standard deviation in weight measurement among the replicated samples. $C$ is a constant to obtain force in Newtons comprised of g = 9.8 ms$^{-2}$ and a conversion factor to change from grams to kilograms.

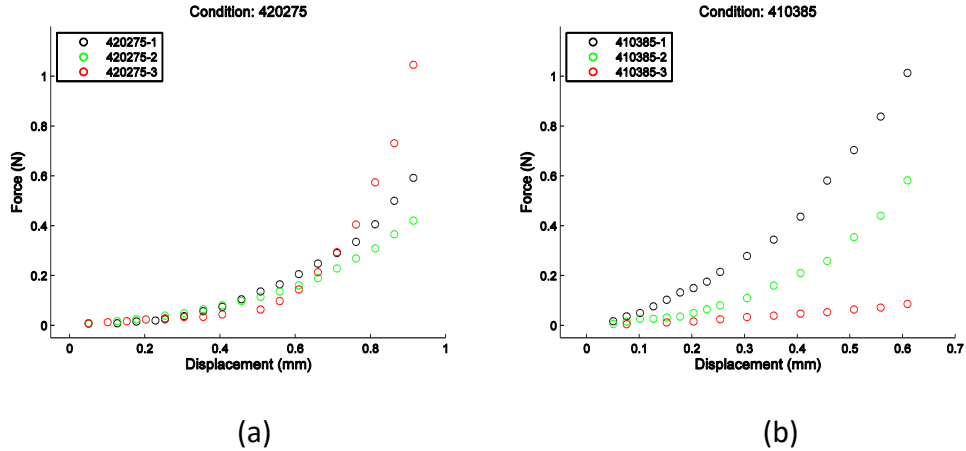

(a) (b)

*Figure 4:* *Force-displacement loading curves of conditions at a.) 75% and b.) 85% relative humidities at 40% w/v nominal PLEY concentration.*

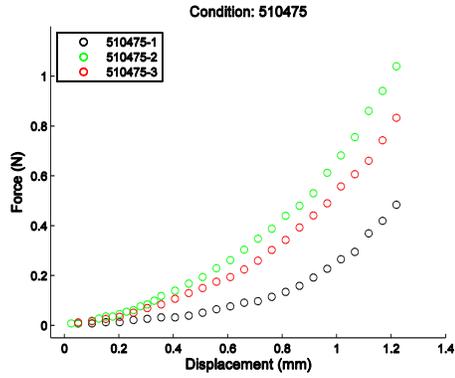 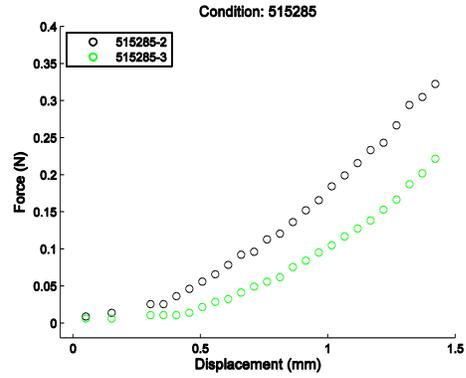

(a) (b)

*Figure 5:* *Force-displacement loading curves of conditions at a.) 75% and b.) 85% relative humidities at 50% w/v nominal PLEY concentration.*

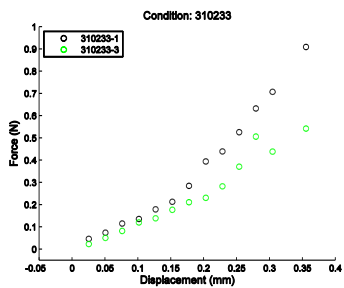 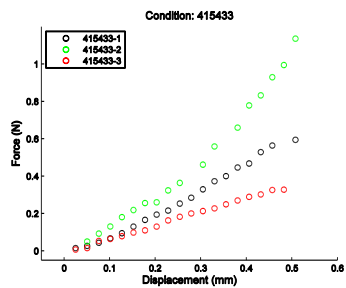 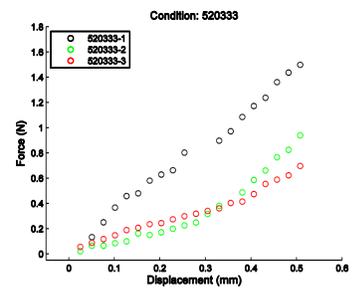

(a) (b) (c)

*Figure 6:* *Force-displacement loading curves of samples at 33% relative humidity with a.) 30% b.) 40% and c.) 50% nominal PLEY concentration.*

Non-linearity was apparent at higher strains for high RH samples as seen in Figs. 3-5.

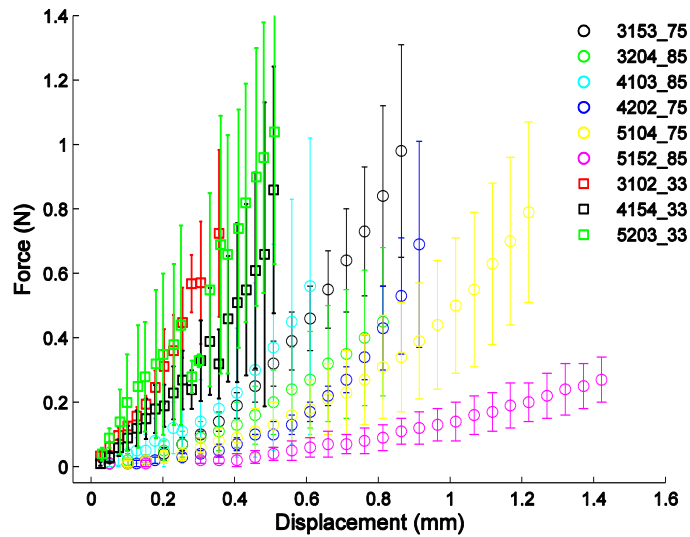

*Figure 7:* *Force-displacement plot with error bars for all conditions presented in Figures 3 - 6. Number after underscore represents relative humidity. For example, relative humidity for 3153_75 is 75% relative humidity. Error in force was calculated as $F = C(w \pm \Delta\varepsilon_w)$, where w is weight and $\Delta\varepsilon_w$ is standard deviation in weight measurement among the replicated samples. $C$ is a constant to obtain force in Newtons.*

Additionally, on samples for which unloading curves could be obtained comparison of loading and unloading curves provided initial evidence of a non-linear elasticity in the present materials. Figure 8 shows force-displacement loading and unloading curves for samples at 75% and 85% RH. Circles represent loading, stars unloading. Loading was clearly non-linear, and unloading closely followed loading. This behavior presented initial indications that the material might have non-linear elasticity. Such behavior alone might not necessarily indicate non-linear elasticity, however. Given the apparent non-linear elasticity exhibited in the force-displacement plots by samples at high RH, we proceeded to further analyze such material through application of phenomenological models detailed in the Nonlinear Elasticity Analysis section.

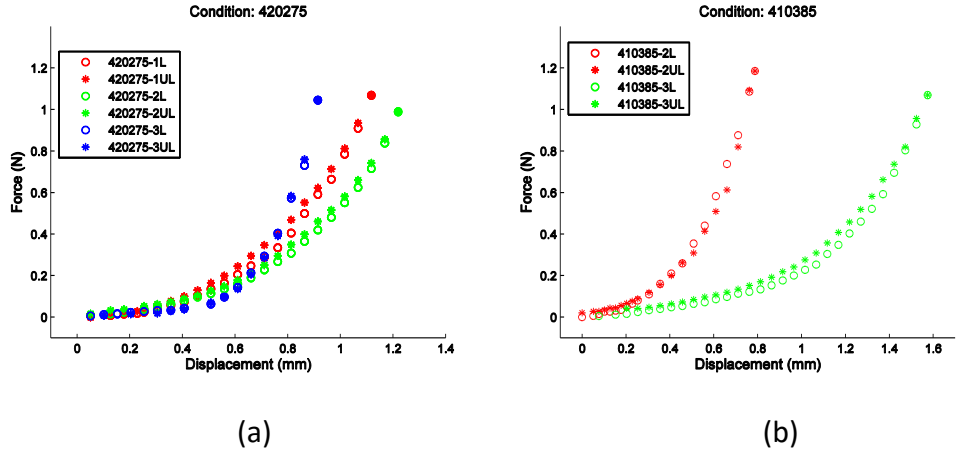

(a)                          (b)

*Figure 8:* Loading/unloading curves were obtained for samples in conditions 420275 and 410385. Specific samples are listed in legend of plot. Circles represent loading; stars, unloading. Unloading data for other conditions could not obtained due material fracture or other factors which prohibited measurement. (a) Condition 420275 at 75% RH. All three samples show pattern of nonlinear elastic deformation. (b) Condition 410385 at 85% RH. Samples show pattern of nonlinear elastic deformation.

Samples at 33% RH were significantly stiffer and more brittle than those at high RH. In both low and high RH, linear regions typically appeared at low strains. Arbitrarily fitting a line to the first six data values of the F-D plots Figures 3-6 gives a rough estimate of the material's spring constant as obtained from the slope of the fitted line. On average, spring constants for materials at each relative humidity were: 290 N/m for 75% RH; 240 N/m for 85% RH and 1340 N/m for 33% RH. Force versus displacement measurements revealed a difference in time-dependent behavior between crosslinked polymers at high RH versus those at low RH. High RH samples immediately reached an equilibrium value in reaction force at a given strain. Those at low RH relaxed over a certain period of time prior to settling on an equilibrium force. For samples at 33% RH at strains up to about 0.04 the resulting force relaxed very quickly to equilibrium, typically less than six seconds. But at strains beginning at around 0.06 and larger the material took progressively longer to relax to equilibrium. To obtain force versus displacement curves, a force reading was obtained after about 10 seconds of relaxation time at each strain. High RH samples showed no evidence of time dependence relaxation within the same 10 second time period. As mentioned above, for each sample at high RH equilibrium force was reached immediately at each imposed strain, within the range of strains in these experiments. Therefore, one can say that the minimum time limit over which samples at high RH exhibited non-linear elasticity was 10 seconds. That is to say that within a 10 second time period, stress at each given strain did not relax down to a different value for samples at high RH.

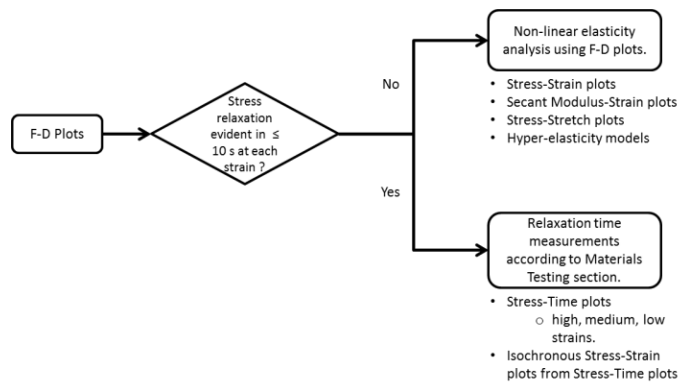

*Figure 9:* *Decision path followed to characterize mechanical properties of crosslinked material in this study.*

Materials at the three low RH conditions clearly exhibited viscoelasticity, consequently, application of hyperelastic models, such as the neo-Hookean or Mooney-Rivlin, were inappropriate to assess elasticity for two reasons. Firstly, the materials exhibited viscoelasticity as opposed to elasticity. The models mentioned above are only applicable to elastic materials. Secondly, stress-stretch plots would show stress values obtained after a 10 second time interval at higher stretches. That would lead to misleading parameter values obtained from such models and not be indicative of actual material behavior. Therefore, our efforts were focused on time-dependent viscoelastic analysis to study relaxation characteristics of materials at low RH as presented in the Nonlinear Viscoelasticity Analysis section.

Figure 9 presents the decision path followed to characterize mechanical properties of crosslinked material in this study. Material at high RH was analyzed for non-linear elasticity since it exhibited no stress relaxation within the range of strains applied and limiting time period used in this study. Material at 33% RH was analyzed as viscoelastic material since it did exhibit stress relaxation at each strain above approximately 0.06.

*Nonlinear Elasticity Analysis*

Many biomaterials display non-linear elasticity.[27] Glandular and fibrous breast tissue under compression, for example, displays mostly linear stress-strain curves below 10% strain; a modulus of 28-35 kPa for glandular tissue and about 96-116 kPa for fibrous tissue at 5% strain, and a modulus of 48-66 kPa and 218-244 kPa at 20% strain.[28] Strain stiffening is also dis-played by artery walls, cornea, blood clots and other biological tissues.[29-31]

To confirm our crosslinked material behaved similar to biological materials, we utilized F-D loading data for analysis of stress-strain curves and looked for evidence of strain hardening in secant modulus versus strain plots. Stress-strain curves were plotted to assess possible non-linear elasticity. Figure 10a shows a typical stress-strain curve. Here it is for condition 515285. Stress-Strain curves for all conditions are presented in Figure 11. To obtain error bars in the stress-strain plot, first stress was calculated as $\sigma = F/A$, where F is the applied force and A is the sample's cross sectional area. Values of force were $F = x_F \pm \Delta\varepsilon_F$, where $x_F$ is each individual value of force at each strain and $\Delta\varepsilon_F$ is variance in force determined from prior calculations.

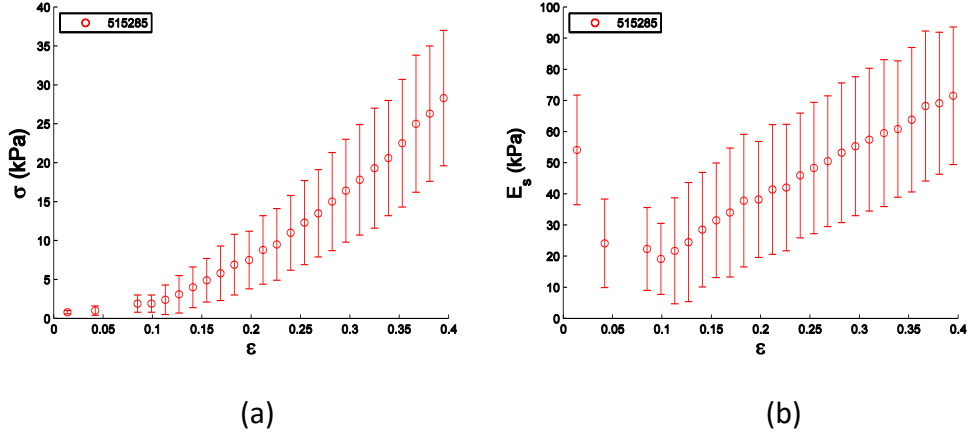

(a) (b)

*Figure 10: Plots of Stress-Strain and Secant Modulus-Strain curves. Strain-Strain curve exhibits signs of strain hardening at high strain. Secant modulus follows a typical pattern of non-linear elastic biological material. Initial decrease in secant modulus, then increased secant modulus with increased strain. (a) Stress versus strain curve for condition 515285; (b) Secant moduli versus strain curve for condition 515285.*

Values of area were $A = x_A \pm \Delta\varepsilon_A$, where the error in area, $\Delta\varepsilon_A$, was calculated as $\Delta\varepsilon_A = A\sqrt{2(\Delta\varepsilon_r/r)^2}$. Average diameter of samples was as given in an earlier section, so that r = 1.75 mm and $\Delta\varepsilon_r$=0.4 mm. The stress with error was then calculated as $\sigma = \frac{F}{A} \pm \Delta\varepsilon_\sigma$, where $\Delta\varepsilon_\sigma = \sqrt{(\Delta\varepsilon_F/F)^2 + (\Delta\varepsilon_A/A)^2}$. It can be seen from the plots that the stress-strain relationship is linear for small deformations. Then, above a threshold value of strain, the relationship becomes non-linear and the material displays strain hardening.

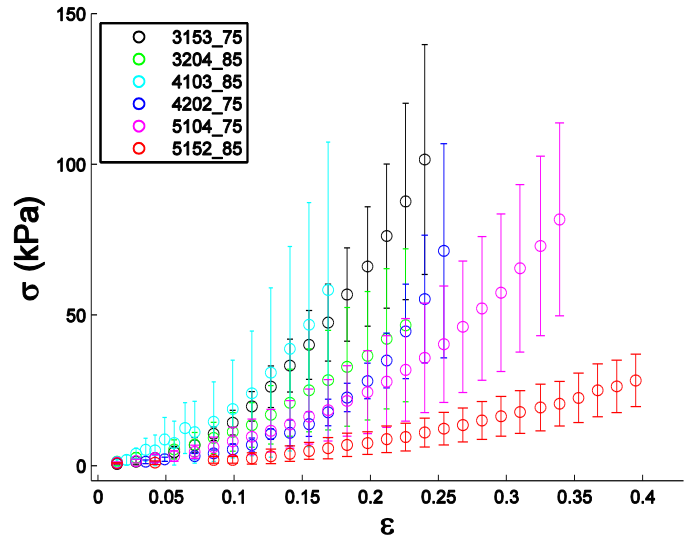

*Figure 11:* Stress versus strain plot with error bars for the six conditions at 75% and 85% RH. Stress was calculated as $\sigma = F/A$, where $F$ is the applied force and $A$ is the sample's cross sectional area. Values of force were $F = x_F \pm \Delta\varepsilon_F$, where $\Delta\varepsilon_F$ is variance in force. Values of area were $A = x_A \pm \Delta\varepsilon_A$, where the error in area, $\Delta\varepsilon_A$, was calculated as $\Delta\varepsilon_A = \sqrt{2(\Delta\varepsilon_r/r)^2}$. Average diameter of samples was $3.5 \pm 0.4$ mm, so that r = 1.75 mm and $\Delta\varepsilon_r$=0.4 mm. The stress with error was then calculated as $\sigma = \frac{F}{A} \pm \Delta\varepsilon_\sigma$, where $\Delta\varepsilon_\sigma = \sqrt{(\Delta\varepsilon_F/F)^2 + (\Delta\varepsilon_A/A)^2}$.

Secant moduli are often used in the analysis of non-linear stress-strain relationships to gain insight on the variation in elasticity as a function of strain.[32-34] The secant modulus is approximately the same as the tangent modulus within the linear regime of small deformations, but it deviates substantially from the tangent modulus in the non-linear region. In the present work, the secant modulus was evaluated at each data point *i* in the stress-strain curve as $(E_s)_i = \frac{\sigma_i}{\varepsilon_i}$, where $\sigma_i$ is engineering stress and $\varepsilon_i$ is engineering strain.

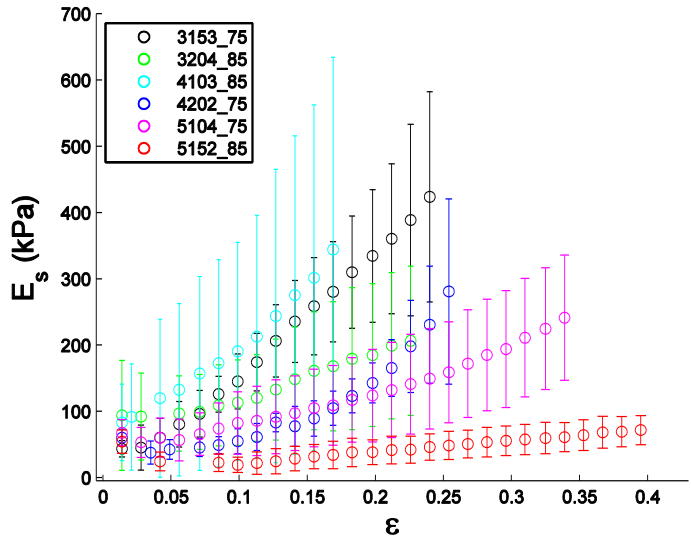

***Figure 12:*** *Secant modulus versus strain plot with error bars for the six conditions at 75% amd 85% RH. Secant modulus was calculated as mentioned in the results and discussion section. We have assumed that $\Delta\varepsilon_\sigma \gg \Delta\varepsilon_\varepsilon$. Therefore the secant modulus with error bars was calculated as $E_s = \frac{1}{\varepsilon}(\sigma \pm \Delta\varepsilon_\sigma)$.*

Figure 10b shows the data of Figure 10a plotted as secant moduli. Typical curves showed an initial decrease in secant modulus at low strain (< 0.05), signaling strain softening. This is probably attributable to polypeptide chain movement and re-ordering upon compressive loading of the material. A further increase of load resulted in strain hardening. Analysis of variation in elasticity, through the secant modulus as a function of strain, provided evidence for strain softening at low strains otherwise not noticeable in the stress-strain plot. Additionally, the analysis pointed to an approximate strain value, different for each condition, where strain softening transitioned to strain hardening. This analysis further substantiated the non-linear elasticity of PLEY material. Figure 12 presents secant moduli for all six conditions. Secant modulus was calculated as mentioned in the results and discussion section. We have assumed that $\Delta\varepsilon_\sigma \gg \Delta\varepsilon_\varepsilon$. Therefore the secant modulus with error bars was calculated as $E_s = \frac{1}{\varepsilon}(\sigma \pm \Delta\varepsilon_\sigma)$, where $\pm\Delta\varepsilon_\sigma$ was calculated previously.

We were interested in calculating the change in secant modulus at each data point, or the approximate second derivative of the stress at each value of strain, as a modified centered-difference second derivative of the stress with respect to strain at each point: $\frac{dE_i}{d\varepsilon}|_{\varepsilon=\varepsilon_i} = \frac{d^2\sigma_i}{d\varepsilon^2}|_{\varepsilon=\varepsilon_i} \approx \frac{\sigma_{i+1}-2\sigma_i+\sigma_{i-1}}{(\varepsilon_i)^2}$. The change in secant modulus, a type of transient response at low to medium strain, was analogous to the second-order transient response to an impulse in an electrical circuit. As strain increased, the change in $E_s$ settled to zero above a certain strain, the value depending on the sample. Figure 13 displays the ratio $\frac{\frac{d^2\sigma_i}{d\varepsilon^2}|_{\varepsilon=\varepsilon_i}}{\frac{d^2\sigma_i}{d\varepsilon^2}|_{\varepsilon=\varepsilon_{max}}}$ at each value of the strain for samples at 75% and 85% RH. One sample from each condition was chosen at random and plotted in this new type of plot for the sole purpose of displaying our results. The interesting behavior, however, calls for additional studies to be conducted using this type of plot. Transient responses are evident up to a strain of $\varepsilon \approx 0.18\text{-}0.2$. For $\varepsilon > 0.2$, $\frac{d^2\sigma_i}{d\varepsilon^2}|_{\varepsilon=\varepsilon_i} \approx 0$. The character of the response appears to support the hypothesis that polypeptide chain movement and re-ordering occur when $\varepsilon \leq 0.18\text{-}0.20$ in materials at 75 and 85% RH. Increased strain could increase polymer alignment and intermolecular interactions. Strain hardening then begins to take place when $\varepsilon > 0.20$ as polymer chains reach an internal entropic force of equal magnitude to that imparted to the material as the polymers elongate between crosslinks. During this stage, the secant modulus fluctuates little if at all, because changes in the elastic modulus between successive strain values are primarily driven by enthalpy rather than chain re-orientation. Consequently, the stress-strain curve becomes non-linear. Fluctuations were greater in general for material at 85% RH than at 75% RH.

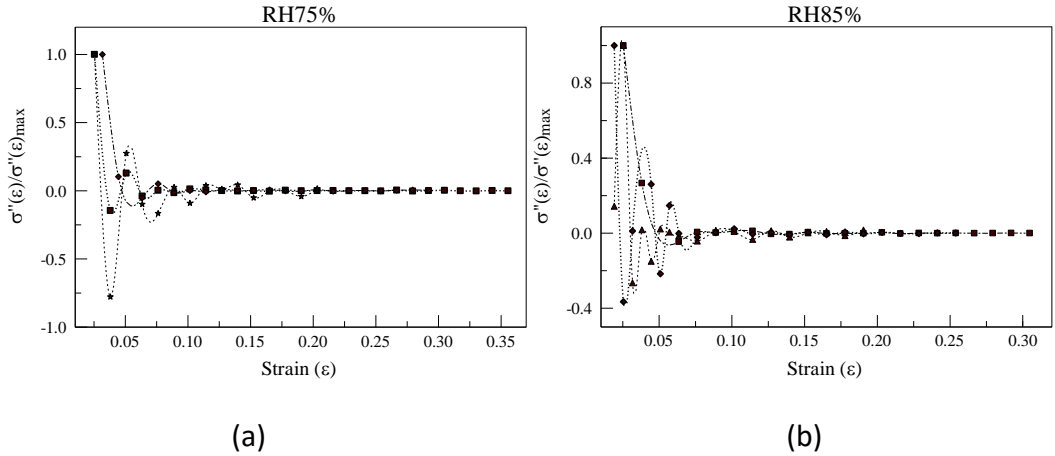

*Figure 13:* Change in secant modulus. (a) Ratio of change in secant modulus at each strain to maximum change in secant modulus versus strain for (a) 75% RH (Samples: 315375-1; 420275-1; 510475-1) and (b) 85% RH (Samples: 320485-1; 410385-2; 515285-2).

Figure 14 compares stress, secant modulus and $d^2\sigma/d\varepsilon^2$ versus strain for condition 515285, sample 515285-3 (85% RH). The data suggest three distinct regions of response to strain. In the first, the secant modulus decreases with increasing strain. This appears to reflect the movement of polypeptide chains relative to each other in a chain re-ordering process. The second derivative displays a transient response in this region. In the second region, there appears to be evidence for strain hardening. The second derivative oscillates around zero with a small amplitude, indicating that the polypeptide chain is still reordering. In the third region, $\frac{d^2\sigma_i}{d\varepsilon^2}\big|_{\varepsilon=\varepsilon_i} \approx 0$. The secant modulus is increasing in response to increasing strain, indicating a continued but moderate rise in stress at each strain in the stress-strain curve. This region corresponds to strain hardening. Similar behavior is displayed by nonlinear elastic biological materials.[33,34] Samples at high RH exhibited non-linear elastic deformation, whereas those at low RH showed non-linear viscoelastic deformation, as will be discussed below.

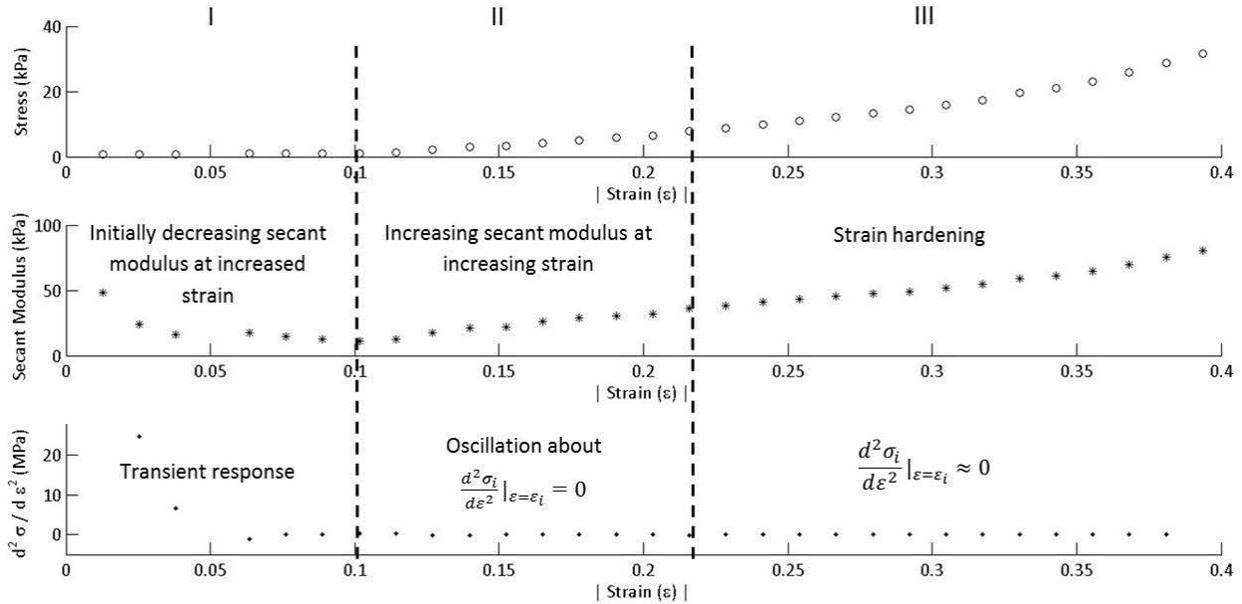

*Figure 14: Nonlinear elastic behavior within each of the three regions in stress-strain response of crosslinked synthetic polypeptide co-poly-(L-glutamic acid$_4$, L-tyrosine$_1$). The data are for condition 515285 using sample 515285-3 to show behavioral characteristics within each of the three regions.*

Figure 15-17 presents stress-stretch data for all conditions at 75% and 85% relative humidities. The applied stress was a uniaxial compression, so the maximum stretch was 1 for $\delta = 0$. Data is represented by the colored open circles. A neo-Hookean model for stress, $\sigma = \mu(\lambda - \frac{1}{\lambda^2})$, was globally fit to the experimental data with $\mu$ as the fitting parameter and represented by a black line. Young's modulus was calculated as $E = 3\mu$ under the assumption of perfect material incompressibility. Table 2 lists shear modulus, $\mu$ from global fits in the second column and calculated Young's Modulus, E, on the third column. The fourth column of table 2 lists the samples at each condition and the fifth column presents shear moduli for each of the samples

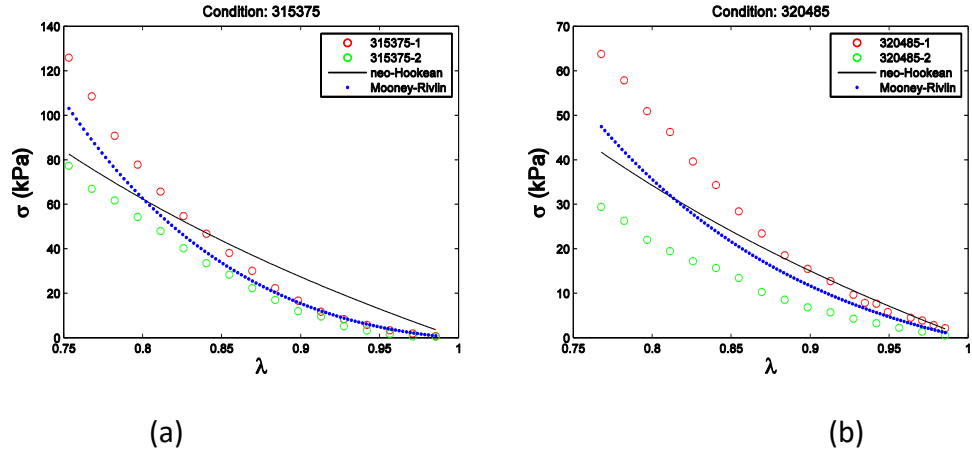

(a)            (b)

*Figure 15:* *Determination of stiffness parameters with global fitting in Matlab®. Circles represent data values; Solid black line is a global fit with a neo-Hookean model. Blue dotted line is a global fit with a Mooney-Rivlin model. Global and individual values of |μ| are listed in Table 2 for the neo-Hookean model and Table 3 for the Mooney-Rivlin model. (a) Stress-stretch curve for condition 315375 (b) Stress-stretch curves for condition 320485.*

(see Supplementary Material for plots of individual fits). The last column calculates the standard deviation of shear moduli among samples within each condition. The sole exception is sample 410385-1, whose data was not included in the global fits of condition 410385.

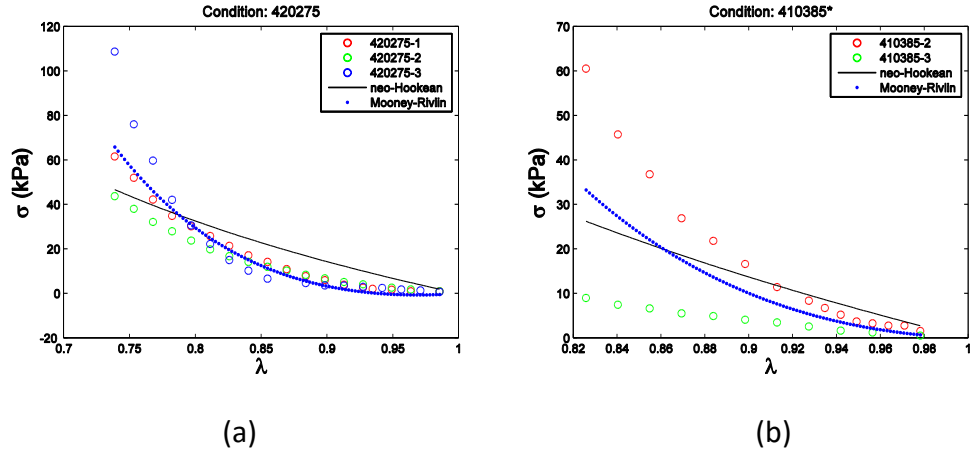

(a)                  (b)

*Figure 16:* *Determination of stiffness parameters with global fitting in Matlab®. Circles represent data values; Solid black line is a global fit with a neo-Hookean model. Blue dotted line is a global fit with a Mooney-Rivlin model. Global and individual values of |μ| are listed in Table 2 for the neo-Hookean model and Table 3 for the Mooney-Rivlin model. (a) Stress-stretch curve for condition 420275 (b) Stress-stretch curves for condition 410385. Note\*: Figure 16(b) excludes sample 410385-1 as data set is an outlier as seen in Tables 2 and 3.*

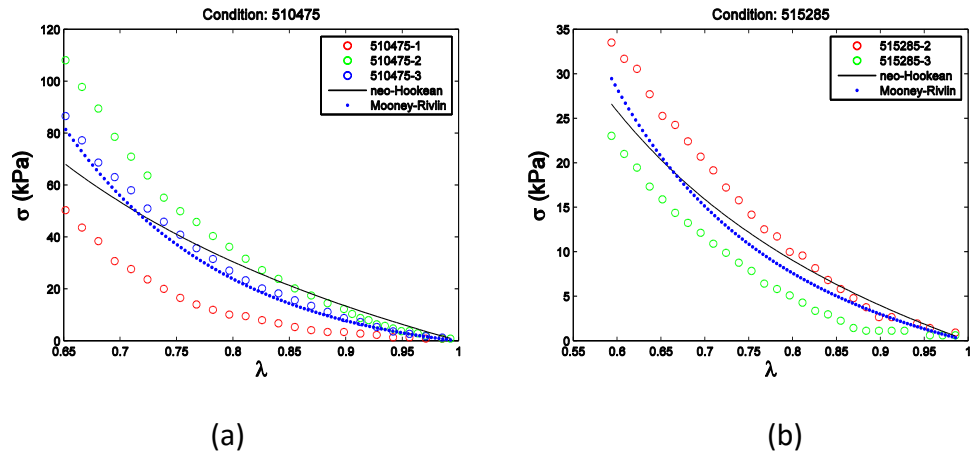

(a)                  (b)

*Figure 17:* *Determination of stiffness parameters with global fitting in Matlab®. Circles represent data values; Solid black line is a global fit with a neo-Hookean model. Blue dotted line is a global fit with a Mooney-Rivlin model. Global and individual values of |μ| are listed in Table 2 for the neo-Hookean model and Table 3 for the Mooney-Rivlin model. (a) Stress-stretch curve for condition 510475 (b) Stress-stretch curves for condition 515285.*

In the same figures we, additionally, present a global fit of data using a two-parameter Mooney-Rivlin model, $= (\mu_1 + \frac{\mu_2}{\lambda})(\lambda - \frac{1}{\lambda^2})$, represented by the blue dotted line. Table 3 lists the fits in the second column. The third and fourth columns show samples at each condition with associated fitting parameters from individual fits (see Supplementary Material for plots of individual fits).

The values shown in the conditions column of table 2 signify the concentrations of PLEY, PLK, EDC and RH. For example, 315375 represents 30% (w/v) PLEY, 15% (w/v) PLK, 30% (w/v) EDC and 75% RH. Data in table 2 shows larger stiffness occurs at PLEY concentrations of 30% compared with 50%. We attribute this to a higher degree of crosslinking for lower PLEY concentrations within the range of EDC concentrations used in this study. Higher concentrations of PLEY might not have resulted in additional crosslinking. Rather, unreacted aqueous PLEY, as well as PLK and EDC, probably remained embedded within the crosslinked network leading to lower elastic moduli. t-ratios (see Figure S3 in supporting information) showed that RH made the most significant contribution to stiffness in the low strain region. The next most influential factors were concentration of PLEY and concentration of PLK. The experiments also revealed that changes in the EDC concentration had little to no effect on the spring constant within the 20-40% range. Polypeptides were covalently crosslinked by EDC but also non-covalently by ionic interactions between PLEY and PLK. A control experiment revealed that a mixture of PLEY and PLK formed a precipitate but the reactants did not crosslink due to the absence of EDC. This indicates that EDC crosslinked PLEY with PLK as expected. 1M EDC was possibly more than sufficient for crosslinking the concentrations used in present experiments so that increased concentrations did not affect crosslinking significantly. The higher values of shear modulus at $\mu$= 139 kPa for sample 410385-1 seems to be an outlier as it doesn't seem to follow a pattern of decreasing relative to conditions at 30% and 50% PLEY. Samples 410385-2 & 3, seem to adhere to the pattern, however. It might be possible that measurement system set-up was somehow slightly different for samples 410385-2 and 410385-3 than that of 410385-1.

Comparing results of $|\mu|$ from global fits of the neo-Hookean model to $|\mu_2|$ from global fits of the Mooney-Rivlin model we see a similar trend. Conditions at 75% RH decrease for increasing PLEY concentration, with $|\mu_2|$=260 kPa for 30% PLEY and $|\mu_2|$=60 kPa at 50%. Again, this was not expected. The trend for values at 85% RH seems less obvious as the condition at 410385 seems to run counter to the trend at 75%. This needs to be investigated further. Regardless, stiffness at 50% RH is lower than stiffness values of either 30% or 40%.

Samples at high RH exhibited non-linear elastic deformation, whereas those at low RH showed non-linear viscoelastic deformation, as will be discussed below.

**Table 2:** Young's moduli values obtained by fitting a neo-Hookean model, $\sigma = \mu(\lambda - \frac{1}{\lambda^2})$, to experimental data in Figures 15-17. First column lists experimental conditions; second lists shear modulus from global fits. E is the Young's modulus. The fourth and fifth columns show samples at each condition with associated shear moduli from individual fits. The last column lists the standard deviation from individual fits (see supplemental information for plots of individual fits).

| Conditions | $|\mu|$ (kPa) | E (kPa) | Sample | $|\mu|$ (kPa) | Std Dev (kPa) |
|---|---|---|---|---|---|
| 315375 | 82 | 246 | 1 | 102 | 24 |
|  |  |  | 2 | 68 |  |
| 320485 | 45 | 135 | 1 | 64 | 25 |
|  |  |  | 2 | 29 |  |
| 420275 | 43 | 129 | 1 | 43 | 12 |
|  |  |  | 2 | 33 |  |
|  |  |  | 3 | 57 |  |
| 410385 | 139 | 417 | 1 | 139 |  |
|  | 41 | 123 | 2 | 70 | 29 |
|  |  |  | 3 | 13 |  |
| 510475 | 40 | 120 | 1 | 22 | 18 |
|  |  |  | 2 | 57 |  |
|  |  |  | 3 | 45 |  |
| 515285 | 12 | 36 | 2 | 15 | 4 |
|  |  |  | 3 | 9 |  |

**Table 3:** Fitting paramaters obtained by fitting the Mooney-Rivlin model, $\sigma = (\mu_1 + \frac{\mu_2}{\lambda})(\lambda - \frac{1}{\lambda^2})$, to experimental data in Figures 15-17. First column lists experimental conditions; second lists shear moduli from global fits. The third and fourth columns show samples at each condition with associated fitting parameters from individual fits (see supplemental information for plots of individual fits).

| Conditions | $[\mu_1, \mu_2]$ (kPa) | Sample | $[\mu_1, \mu_2]$ (kPa) |
|---|---|---|---|
| 315375 | [240, -260] | 1 | [360, -370] |
|  |  | 2 | [170, -190] |
| 320485 | [60, -90] | 1 | [100, -130] |
|  |  | 2 | [50, -60] |
| 420275 | [220, -210] | 1 | [170, -170] |
|  |  | 2 | [90, -100] |
|  |  | 3 | [470, -420] |
| 410385 | [540, -590] | 1 | [540, -590] |
|  | [210, -220] | 2 | [460, -460] |
|  |  | 3 | [16, -25] |
| 510475 | [40, -60] | 1 | [60, -60] |
|  |  | 2 | [50, -80] |
|  |  | 3 | [40, -60] |
| 515285 | [-1, -7] | 2 | [-6, -6] |
|  |  | 3 | [6, -11] |

*Nonlinear Viscoelasticity Analysis*

Viscoelastic materials play an important role in applications requiring energy absorption. Examples include earthquake dampers and cushioning in seats and shoes. Such materials could benefit from engineering viscoelastic properties.  Viscoelasticity is an important feature of biological materials.  The time dependence of stress-strain relationships could potentially be engineered to meet the requirements of medical or non-biological applications of polypeptide materials.  The present materials and others based on designed polypeptides have potential applications in cartilage replacement.  We therefore determined certain viscoelastic properties of the present polypeptide materials.

Samples at 33% RH showed a viscoelastic response under compression.  At each increment of displacement from equilibrium (strain), time was required for the stress measurement to come to equilibrium, the amount depending on the sample.  In general, samples at 75% RH and 85% RH did not display such behavior, the sole exception being Sample 515285-2, which was compressed to a relatively large deformation.  Due to the observed viscoelastic response, relaxation times for samples at RH33% were therefore obtained.

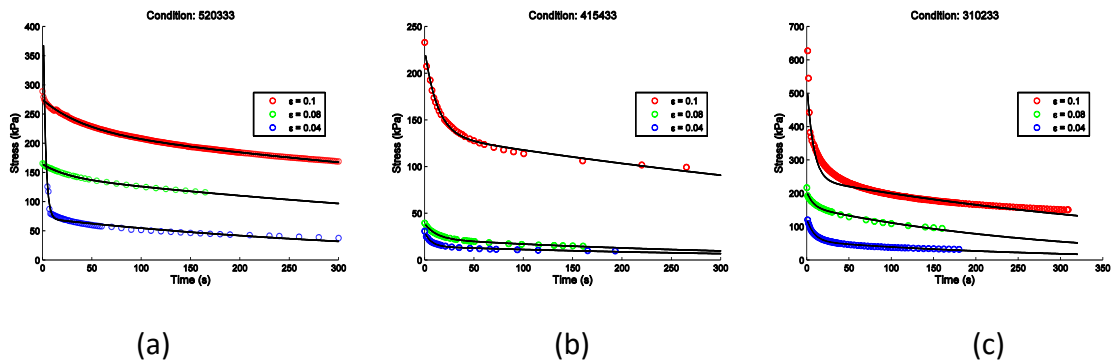

(a)                        (b)                        (c)

*Figure 18: Relaxation time data for samples 310233-1; 415433-2 and 520333-1 at 33% relative humidity and at high, medium and low strains. High strain corresponds to approximately 10% deformation, medium strain to 8% deformation and low strain to 4% deformation. Colored circles represent data; solid line represents the fit. Only relaxation times were important for this study. (a) Relaxation time data for sample 310233-1. (b) Relaxation time data for sample 415433-2. (c) Relaxation time data for sample 520333-1.*

There were three conditions for material at 33% RH of three samples each for a total of nine samples. Due to material brittleness some of the samples that survived processing did not survive F-D measurements. However, samples that survived processing and F-D measurements were used to study relaxation characteristics. It was hypothesized that samples available for time relaxation studies would not behave much different, in general, than other samples within each respective condition. This hypothesis can be disproved in future studies. Measurements on available samples were, thus, obtained to study characteristic relaxation behavior under low, medium and high strains.

To obtain relaxation times, we applied a predetermined strain at a low value, where it was assumed that Hooke's law applied, at a medium value, and a high value, well above the probable elastic region. In general, the first eight data points of samples at 33% RH extended to displacements of $\delta = 0.2$ mm, and stress showed a linear response to strain. For $\delta > 0.2$ mm, stress-strain relations became nonlinear up to the maximum displacements measured in this study, $\delta \approx 0.7$ mm. The high strain was obtained by displacing samples by 0.41 mm, a compressive deformation of c. 10%. This enabled sample analysis at a point close to the upper end of the non-linear region. Medium strain was at a displacement of 0.30 mm, a c. 8% deformation. Low strain was obtained by displacing the sample by 0.15 mm, a c. 4% deformation.

Figure 18 shows relaxation data for 310233-1; 415433-2 and 520333-1 at high, medium and low strain. Red symbols are for 10% deformation, green for 8% deformation, and blue for 4% deformation. Figure 18b shows the fit $\sigma(t) = \varepsilon_o \left( E_1 e^{-\frac{t}{\tau_1}} + E_2 e^{-\frac{t}{\tau_2}} \right)$ to experimental low-strain data. Table 4 presents relaxation times of selected samples as obtained by fitting $\sigma(t) = \varepsilon_o \left( E_1 e^{-\frac{t}{\tau_1}} + E_2 e^{-\frac{t}{\tau_2}} \right)$ to experimental data sets. Shorter values of $\tau_1$ can be attributed to the fast relaxation due to the spring components of the Maxwell system, larger values of $\tau_2$ are attributable to damping of the dashpot, or viscous, components. For example, Figure 18b shows a steep decrease in stress within the first 13 s of strain at $\varepsilon = 0.04$. This fast relaxation could be attributable to the effect of short-order polymer network re-alignment in response to the step increase in stress, typically modeled as a Heaviside step function. Further network re-alignment is slowed down by the viscous component of surrounding medium, which results in longer-term stress relaxation. Figure 18b shows such gradual decrease in stress for times greater than 13 s.

*Table 4: Relaxation times for three representative samples at 33% RH and at high, medium or low strains.*

|  |  | Relaxation time | | | | | |
|---|---|---|---|---|---|---|---|
|  |  | High Strain ($\varepsilon$ = 0.10) | | Medium Strain ($\varepsilon$ = 0.08) | | Low Strain ($\varepsilon$ = 0.04) | |
| Condition | Sample | $\tau_1$ (s) | $\tau_2$ (s) | $\tau_1$ (s) | $\tau_2$ (s) | $\tau_1$ (s) | $\tau_2$ (s) |
| 310233 | 310233-1 | 7.6 | 470 | 7.5 | 290 | 8.9 | 280 |
| 415433 | 415433-2 | 13 | 770 | 12 | 360 | 6.9 | 410 |
| 520333 | 520333-1 | 43 | 1100 | 25 | 770 | 2.4 | 380 |

Figure 19 presents isochronous stress-strain plot.[35] A nonlinear isochronous stress-strain plot signifies nonlinear viscoelasticity. The samples were at 33% RH. The strain values were as in Table 4, and data were compared at three different time points: 10 s, 100 s and 150 s. Relaxation plots similar to Figure 18b were obtained for each value of the strain. For example, Figure 18b shows relaxation data for sample 415433-2 at low strain where at 10 s, σ = 18 kPa, at 100 s, σ = 11 kPa, and at 150 s, σ = 10 kPa. These data are plotted for ε = 0.04 in Figure 19b. The same process gave points for ε = 0.08 and ε = 0.10. The upper curve in each panel of Figure 19 is for 10 s, the middle curve, 100 s, and the lower curve, 150 s. It is clear that these 33% RH samples behaved as nonlinear viscoelastic materials. Materials at 33% RH showed obvious viscoelastic behavior throughout the range of applied strain.

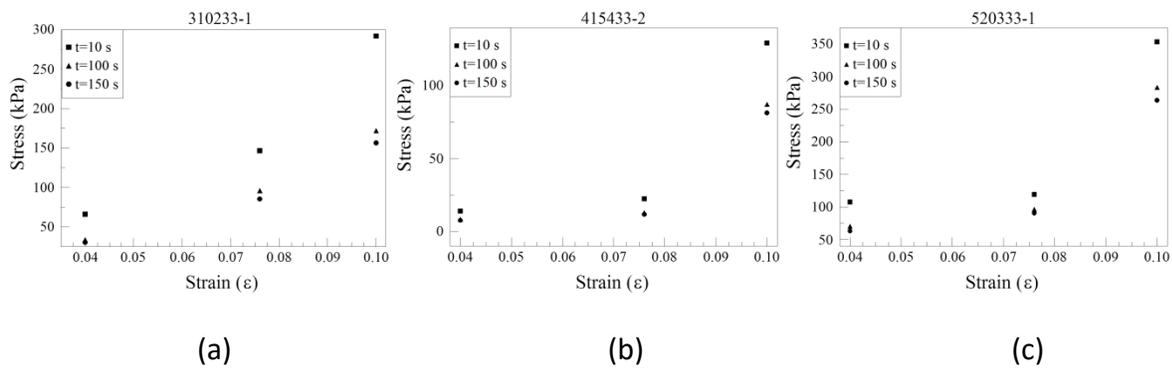

(a)  (b)  (c)

*Figure 19:* Isochronous stress-strain plots of samples in three different factors at 33% RH. The three lines represent 10 s, 100 s and 150 s in each case. (a) Sample 310233-1,(b) Sample 415433-2, (c) Sample 520333-1. Nonlinearity provides evidence of nonlinear viscoelasticity.

Materials at 75% or 85% RH could potentially display viscoelasticity beyond a strain threshold. This possibility is supported by analysis of sample 515285-2. At 85% RH, the sample exhibited viscoelastic behavior at displacements beyond 0.86 mm, or 22% deformation on a sample height of 4 mm. At displacements above 0.86 mm viscoelastic behavior was found. Single-exponential relaxation times were found to be 1400 s and 650 s for high and low stress, respectively.

**CONCLUSION**

Post-processing of data revealed significant variability among replicate samples within each condition. Variability among samples and conditions stemmed from processing and measurement methods. First, reactant mixtures were made to nominal concentrations, each with approximately 2% uncertainty due to weighing of lyophilized polymer and addition of solvent water with micro-pipettes. Transfer of reactants to test tube proved to be a difficult task at high PLEY concentrations due its high viscosity. Mixture of reactants resulted in approximately 5% uncertainty a previously stated. Transfer of reactant mixture at nominally 120 µL to the mold with a syringe resulted in additional losses of about five to six microliters, or 5% uncertainty. Relative humidity chambers are expected to be within 2% of nominal. However, typically four to five samples were placed in their respective relative humidity chambers as called for in Table S1. Not all samples in each RH chamber were measured at once due to a strict adherence to conduct measurements down the list shown in Table S1. Therefore, some samples experienced cycles of changing RH as chambers were opened and closed to measure samples that were chronologically next in line. Due to this changing environment, it is estimated that samples experienced relative humidities within a 4% uncertainty. The sum total of these uncertainties is 20%. Additionally, care had to be taken during measurement to ensure the fixed arm was making proper contact with the sample. Improper con-tact between fixed arm and sample created immediately noticeable errors in readings. It was therefore critical that proper contact be made at each measurement run. Resolution of scale was within 0.1 g and the displacement gage was within 0.025 mm. It is possible that the measurement apparatus as a whole provided approximately 10% uncertainty. Plots with error bars clearly show that higher variability occurred at high strains. Most samples within each condition seem to follow similar patterns at low displacements (strains), with the exception of condition 410385. This seems to indicate that a majority of sample variability arose from sample fabrication. The sum total of fabrication and measurement is about 30% uncertainty.

Using the coefficient of variation, CV = (Std Dev)/(mean $|\mu|$), to estimate data value dispersion we find uncertainty in shear modulus, obtained from the neo-Hookean model, to be the following for each condition: $CV_{315375}$ = 30% ; $CV_{320485}$ = 56% ; $CV_{420275}$ = 28% ; $CV_{410385}$ = 71% ; $CV_{510475}$ = 45%; $CV_{515785}$ = 33%. Other than the two conditions where CV > 50% the rest seem to be close to expectation.

Of particular importance is the uncertainty with regards to relative humidity. Due to the 4% uncertainty in RH it is difficult to distinguish and compare stiffness properties between conditions at 75% and 85%. However, within the 75% RH condition a trend of decreasing stiffness with increasing PLEY concentration is readily apparent. This trend is not so obvious at 85% RH since condition 410385 does not follow the same pattern. Excluding the first sample as an outlier, shows that condition 410385 is similar in stiffness as 320485. This could be either due to errors in processing, measurement, difference in PLK concentration or a combination of all three. It could also be possible that measurements are correct and represent actual material behavior. This should be investigated further in future studies.

Due to sample variability in this study, values for stiffness here are to be understood as for comparative purposes only to ascertain trends as they relate to changes in conditions.

As far as trends are concerned, we conclude that the mechanical properties exhibited by cross-linked PLEY are typical of biomaterials. Loading/unloading curves revealed that the present polypeptide materials dis-played nonlinear viscoelasticity at low relative humidity (33%). This was in contrast to the display of nonlinear elasticity at high relative humidity (75-85%). At low strains, the material shows low elasticity as polypeptide chains move relative to each other. As strain increases beyond a certain point, the material shows sign of strain hardening as the secant modulus increases. The material seems to behave close to a neo-Hookean model from solid mechanics. It is presumed that water inclusion within polypeptide cross-linked structures gives rise to this behavior. However, more testing needs to be conducted.

Further, we conclude that out of the four factors studied here, the most significant with regards to material stiffness was relative humidity. Higher relative humidities result in lower stiffness. Lower relative humidities result in higher stiffness. This is ascribed to the higher load bearing capability of cross-linked polypeptides primarily comprised of covalent bonds that are comparatively stronger than the hydrogen or van der Waals bonds manifested by water molecules at higher relative humidites.

Lastly, it was shown that cross-linked polypeptide material at low humidities exhibit relaxation times on the order of minutes. For those studied in this paper, at high deformation viscosity-driven relaxation times averaged 13 min. At low deformation, viscosity-driven relaxation times averaged 6 min.

We hope this work opens new areas of research into the engineering of designed polypeptides to obtain specifically desired relaxation times for biomedical applications or applications where biodegradability is desired.